\documentstyle[aps,pra]{revtex}  
\begin{document}                
\draft
\title{\bf A proposal of quantum logic gates using cold trapped ions in a 
cavity}
\author{F.L. Semi\~ao\footnote{e-mail: semiao@ifi.unicamp.br}, A. 
Vidiella-Barranco\footnote{e-mail: vidiella@ifi.unicamp.br} and J.A. 
Roversi\footnote{e-mail: roversi@ifi.unicamp.br}}  
\address{Instituto de F\'\i sica ``Gleb Wataghin'',
Universidade Estadual de Campinas,
13083-970   Campinas  SP  Brazil}  
\date{\today}
\maketitle
\begin{abstract}    
We propose a scheme for implementation of logical gates in a trapped ion
inside a high-Q cavity. The ion is simultaneously interacting with a (classical) 
laser field as well as with the (quantized) cavity field. We demonstrate 
that simply by tuning the ionic internal levels with the frequencies of the fields,  
it is possible to construct a controlled-NOT gate in a three step procedure, having 
the ion's internal levels as well as vibrational (motional) levels as qubits. The 
cavity field is used as an auxiliary qubit and basically remains in the vacuum 
state.
\end{abstract}

The coherent manipulation of simple quantum systems has become increasingly 
important for both the fundamental physics involved and prospective applications, 
especially on quantum information processing. Entanglement between two or more
subsystems is normally required in order to have conditions for ``quantum logical''
operations to be performed. Two-level systems are natural candidates for building 
quantum bits (qubits), which are the elementary units for quantum information 
processing.  
We may mention single ions interacting with laser fields \cite{wine0}, atoms and field 
modes inside high-Q cavities \cite{haroche0}, and molecules (via NMR) \cite{chuang},
as quantum subsystems which have shown themselves suitable for coherent manipulation. 
Regarding the atoms (or ions), both internal (electronic) as well as vibrational motion 
states may be readily used for performing quantum operations, e.g., a controlled-NOT 
gate \cite{wine1}, and a phase gate \cite{haroche0}. It is therefore important to 
explore other combinations of (experimentally available) physical systems. An 
interesting set up is a single trapped ion inside a cavity. 
The quantized field couples to the oscillating ion so that we have three quantum 
subsystems: the center-of-mass ionic oscillation, the ion's internal
degrees of freedom, and the cavity field mode. One of the advantages of such a system is
the high degree of control one may achieve in trapped ions, allowing, for instance, long
interaction times with the cavity field. A few papers may be found, in which it
is investigated the influence of the field statistics on the ion dynamics 
\cite{zeng,knight}, quantum state transfer \cite{parkins}, 
as well as a scheme to generate Bell-type states of the cavity-field
and the vibrational motion \cite{ours0}. More recently, we may find propositions of other
schemes involving the combination of trapped ions with cavity QED \cite{walther1,jane}.
On the experimental side, a single trapped ion has been succesfully coupled to a cavity 
field \cite{walther,blatt}, an important step towards the use of trapped ions for 
quantum computing and quantum communication purposes \cite{steane}.   

In this paper we present a new scheme for quantum state manipulation from which we may
construct a quantum phase gate and a Hadamard gate, making possible to implement a 
controlled-NOT gate. This could be accomplished simply by adjusting atom-field 
detunings within the same experimental set-up. A controlled-NOT gate transforms 
two-qubit states in the following way
\begin{eqnarray}
&&|0,0\rangle \Longrightarrow |0,0\rangle \nonumber\\ 
&&|0,1\rangle \Longrightarrow |0,1\rangle \nonumber\\ 
&&|1,0\rangle \Longrightarrow |1,1\rangle \nonumber\\ 
&&|1,1\rangle \Longrightarrow |1,0\rangle,
\end{eqnarray}
i.e., the second qubit (target qubit) undergoes a change only if the first qubit
(control qubit) is one, no changes occurring if its value is zero. Our scheme is 
based on a single ion trapped inside a cavity, having the ion coupled to both the 
cavity field as 
well as to a classical driving field. We also need an auxiliary state in order to 
perform the operations above. For instance, in the method presented in reference 
\cite{cirac0}, other internal atomic states may be used as auxiliary qubits.
In our scheme, the two internal ionic levels represent the target qubit, and the 
ionic center-of-mass oscillation is the control qubit. Differently from other
schemes, though, here the {\it quantized field states} constitute the  auxiliary qubit 
necessary for the implementation of a controlled-NOT logic gate. 

It is not difficult to show that a controlled-NOT gate is equivalent to the 
application of a Hadamard gate, followed by the application of a phase gate,
and an application of a Hadamard gate again.
A Hadamard gate has the following action
\begin{eqnarray}
\label{hadamard}
&&|0\rangle \Longrightarrow \frac{|0\rangle + |1\rangle}{\sqrt{2}} \nonumber \\ 
&&|1\rangle \Longrightarrow \frac{|0\rangle - |1\rangle}{\sqrt{2}},
\end{eqnarray}
and a phase gate (causing a particular phase shift of $\pi$) is such that
\begin{eqnarray}
\label{phasegate}
&&|0,0\rangle \Longrightarrow |0,0\rangle \nonumber \\ 
&&|0,1\rangle \Longrightarrow |0,1\rangle \nonumber \\ 
&&|1,0\rangle \Longrightarrow |1,0\rangle \nonumber \\ 
&&|1,1\rangle \Longrightarrow -|1,1\rangle.
\end{eqnarray}
We shall seek for convenient interactions which could allow the implementation of 
the sequence of operations described above, bearing in mind the importance of the 
speed of operation of the logic gate, specially because of the unwanted action of
the environement that normally causes decoherence.

We consider the interaction of a (two-level) trapped ion in interaction with an
external laser field (having frequency $\omega_L$), as well as with a cavity field 
(having frequency $\omega_c$). The hamiltonian for such a system reads
\begin{eqnarray}
\hat{H}&=&\hbar\nu\hat{a}^\dagger\hat{a}+\hbar\omega_c \hat{b}^\dagger\hat{b}+ 
\frac{\hbar\omega_0}{2}\sigma_z + 
\hbar\Omega\sigma_+ \exp\left[i\eta_L(\hat{a}^\dagger+\hat{a})-i\omega_L t\right]+ 
\nonumber \\ 
&+&\hbar\Omega\sigma_- \exp\left[-i\eta_L(\hat{a}^\dagger+\hat{a})+i\omega_L t\right]+ 
\hbar g (\sigma_+ + \sigma_-)(\hat{b}^\dagger+\hat{b})
\sin\eta_c(\hat{a}^\dagger+\hat{a}),
\end{eqnarray}
where $\hat{a}^\dagger (\hat{a})$ are the creation (annihilation) operators relative to
the vibrational motion excitations, $\hat{b}^\dagger (\hat{b})$ are the creation 
(annihilation) operators of the cavity field excitations, $\sigma_+ (\sigma_-)$ are the
raising (lowering) atomic operators, $\omega_0$ is the atomic frequency, $\nu$ is the
ionic vibrational frequency, and $\eta_L$, $\eta_c$ are the Lamb-Dicke parameters
relative to the laser field and the cavity field, respectively. We assume a ``double'' 
Lamb-Dicke regime, or $\eta_L\ll 1$ and  $\eta_c\ll 1$, so
that we may write $\exp\left[i\eta_L(\hat{a}^\dagger+\hat{a})\right]\approx 1+
i\eta_L(\hat{a}^\dagger+\hat{a})$ and $\sin\eta_c(\hat{a}^\dagger+\hat{a})\approx
\eta_c(\hat{a}^\dagger+\hat{a})$. Under that approximation, the interaction hamiltonian
in the interaction picture will be
\begin{eqnarray}
\label{fullhamil}
\hat{H}_i&=&  
i\eta_L\hbar\Omega\left[\sigma_+ \hat{a}\exp\{ i(\delta_{aL}-\nu) t\} - h.c.\right] +
\eta_c \hbar g\left[\sigma_+ \hat{a}^\dagger\hat{b}
\exp\{ i(\delta_{ac}+\nu) t\} + h.c.\right]
\nonumber \\ 
&+&i\eta_L \hbar\Omega\left[\sigma_+ \hat{a}^\dagger
\exp\{ i(\delta_{aL}+\nu) t\}-h.c.\right] + 
\eta_c \hbar g\left[\sigma_+ \hat{a}^\dagger\hat{b}^\dagger
\exp\{ i(\delta_{ac}+\nu + 2 \omega_c) t\} + h.c.\right] + \nonumber \\ 
&+&\eta_c \hbar g\left[\sigma_+ \hat{a}\hat{b}^\dagger
\exp\{ i(\delta_{ac}-\nu + 2 \omega_c) t\} + h.c.\right] + 
\eta_c \hbar g\left[\sigma_+ \hat{a}\hat{b}
\exp\{ i(\delta_{ac}-\nu) t\} + h.c.\right] + \nonumber \\ 
&+&\hbar\Omega[\sigma_+ \exp(i(\delta_{aL} t) + h.c.] \nonumber
\end{eqnarray}
where $\delta_{aL}=\omega_0-\omega_L$ and $\delta_{ac}=\omega_0-\omega_c$.

In order to implement the controlled-NOT gate in the system considered here, it is
required the following: for the Hadamard gate [see Eq. (\ref{hadamard})], we need 
an interaction hamiltonian of the type
\begin{equation}
\label{hadhamil}
\hat{H}_{Hg}=\hbar\Omega [\sigma_+ + \sigma_-],
\end{equation}
and an application of a $\pi/2$ laser pulse. 
For the phase gate [see Eq. (\ref{phasegate})], it is needed the following
interaction hamiltonian 
\begin{equation}
\label{phahamil}
\hat{H}_{pg}=\hbar\eta g [\sigma_+\hat{a}^\dagger\hat{b} + \sigma_-\hat{a}\hat{b}
^\dagger],
\end{equation}
and an application of a $2\pi$ pulse.

For instance, the action of the phase gate above is such that 
\begin{eqnarray}
\label{action}
\left[|0\rangle_v |g\rangle\right] |0\rangle_f & \longrightarrow &
\left[|0\rangle_v |g\rangle\right] |0\rangle_f \nonumber \\ 
\left[|0\rangle_v |e\rangle\right] |0\rangle_f & \longrightarrow &
\left[|0\rangle_v |e\rangle\right] |0\rangle_f \nonumber \\ 
\left[|1\rangle_v |g\rangle\right] |0\rangle_f & \longrightarrow &
\left[|1 \rangle_v |g\rangle\right] |0\rangle_f \nonumber \\ 
\left[|1\rangle_v |e\rangle\right] |0\rangle_f & \longrightarrow &
-\left[|1 \rangle_v |e\rangle\right] |0\rangle_f. 
\end{eqnarray}
It is simply required a cavity field (auxiliary qubit) initially prepared in the vacuum 
state. After an operation takes place, (e.g., after a $2\pi$ pulse) the cavity field 
remains in the vacuum state, i.e., it is left ready for newcoming operations, as we see
in (\ref{action}). The system described by the hamiltonian in Eq. 
(\ref{fullhamil}) makes possible the implementation of a Hadamard gate and a phase gate 
simply by tuning the atomic levels relatively to the cavity and laser 
fields. If $\delta_{aL}=0$ (or $\omega_0=\omega_L$) in (\ref{fullhamil}) (after 
applying the rotating wave approximation),  we end up with the hamiltonian necessary 
for the implementation of a Hadamard gate [see Eq. (\ref{hadhamil})]. On the other 
hand, if $\delta_{ac}=\omega_0-\omega_c=-\nu$, the obtained hamiltonian will be 
precisely the one needed for the phase gate operation [see Eq. (\ref{phahamil})]. 
Since it is necessary to perform a Hadamard gate plus a phase 
gate and one Hadamard gate again to have a controlled-NOT gate, there is need of
rapidly switching from the interaction hamiltonian in Eq. (\ref{hadhamil}) 
(after applying a $\pi/2$ laser pulse), to the one in Eq. (\ref{phahamil}) and back to 
(\ref{hadhamil}). This may be accomplished by applying static electric fields 
in order to tune (via Stark effect, for instance) the atomic energy levels either with 
the laser field or the cavity field.

A question that normally arises is about the destructive effect of decoherence on the 
gate operation, which represents one of the main obstacles for the implementation of 
such schemes. In experiments using a trapped ion inside a cavity 
performed so far \cite{walther,blatt}, the cavity fields are in the optical regime, and 
the presently available optical cavities \cite{parkins} have a decay time 
$\tau_c\approx 1\, \mu$s. 
Here we have sources of decoherence from both the trapped ion as well 
as from cavity losses. Heating of the ion is one of the causes of decoherence in 
the harmonic motion, even though coherence may be maintained for times as long as 1 ms in 
current experiments \cite{blatt0}. Regarding the (internal levels) spontaneous decay, 
one way of avoiding it is by coupling non-dipole allowed transitions via lasers in a 
Raman configuration, for instance \cite{jane}. Another possibility would be to make use of
long-lived states (around 1 s) in ${}^{40}\mbox{Ca}^{+}$, as discussed in \cite{blatt1}.
In a recent experiment, the quadrupole S$_{1/2}$ - D$_{5/2}$ transition at 729 
nm has been succesfully coupled to a cavity field \cite{blatt}. Although that is a 
relatively weak coupling, it represents an important step towards the coherent 
control required for quantum information processing.
In our method, it would take at least a time $\tau_{Had}=\pi/4\Omega$ to perform a 
Hadamard gate and $\tau_{pha}=\pi/\eta g$ to perform a phase gate. 
We estimate\footnote{We have used $\Omega=g\approx 2\pi\times 10^5$ Hz for the
optical case.} that a 
minimum time $\tau_{opt}\approx 5\, \mu$s is required for the three step procedure in 
the optical case, which is in fact longer than the cavity decay time in the optical 
range. This means that we need improvements in the available optical technology, 
such as higher finesse cavities, for the implementation of the scheme presented above.
On the other hand, in the microwave regime the photon lifetime is much longer, around 
$\tau_c\approx 0.2$ s \cite{walther2}, and it is needed a time 
$\tau_{mic}\approx 30\,\mu$s for the gate operation. However, similar experiments 
have not been realized in the microwave domain. 
We would like to remark that the auxiliary qubit (cavity field) is
basically left in the vacuum state, which enhances the robustness of our scheme 
against dissipation. Another advantage of our scheme is that control is achieved in
a relatively simple way through the application of external fields in a three step
sequence.

We have presented an alternative scheme that would allow the implementation of quantum 
logical operations in cold trapped ions. The ion is supposed to be placed
inside a high finesse cavity so that the gate operation is assisted by the (quantized)
cavity field. A controlled-NOT gate may be constructed by applying laser pulses during
convenient times, and also tuning the ion's internal levels either with the cavity 
field or with an external (classical) field. The cavity field remains most of the time
in its vacuum state during the logical operations, which is desirable if one wants to
avoid the destructive effects of cavity losses. We conclude that trapped ions $+$ 
cavities represent a very promising system for quantum information processing.

{\bf{Acknowledgments}}

This work is partially supported by CNPq (Conselho Nacional para o 
Desenvolvimento Cient\'\i fico e Tecnol\'ogico), and FAPESP (Funda\c c\~ao 
de Amparo \`a Pesquisa do Estado de S\~ao Paulo), Brazil, and it is linked 
to the Optics and Photonics Research Center (CePOF, FAPESP).

\end{document}